\documentclass[11pt]{elsarticle}
\usepackage{color,soul}
\usepackage{booktabs}
\usepackage[normalem]{ulem}
\usepackage{subcaption}
\useunder{\uline}{\ul}{}
\usepackage{notoccite}
\usepackage{float}
\usepackage{enumitem}
\usepackage{natbib}
\setcitestyle{authoryear,open={(},close={)}}
\usepackage{soul}
\usepackage{url}
\usepackage{algorithm2e}
\usepackage{amsmath,bm}
\usepackage{xcolor}

\usepackage{paralist}

\setlist[enumerate]{nosep}
\usepackage{graphicx}
\usepackage{amssymb}
\usepackage{lineno}

\usepackage{geometry}
\journal{Computers, Environment and Urban Systems}

\date{2024}
\begin{document}

\title{GeoAI-Enhanced Community Detection on Spatial Networks with Graph Deep Learning}

\author[add1]{Yunlei Liang }
\author[add1,add2]{Jiawei Zhu }
\author[add1,add3]{Wen Ye}
\author[add1]{Song Gao*}\ead{song.gao@wisc.edu}
 
\address[add1]{GeoDS Lab, Department of Geography, University of Wisconsin-Madison}
\address[add2]{School of Architecture and Art, Central South University}
\address[add3]{Department of Computer Science, University of Southern California}

\begin{abstract}
Spatial networks are useful for modeling geographic phenomena where spatial interaction plays an important role. To analyze the spatial networks and their internal structures, graph-based methods such as community detection have been widely used. Community detection aims to extract strongly connected components from the network and reveal the hidden relationships between nodes, but they usually do not involve the attribute information. To consider edge-based interactions and node attributes together, this study proposed a family of GeoAI-enhanced unsupervised community detection methods called \textit{region2vec} based on Graph Attention Networks (GAT) and Graph Convolutional Networks (GCN). The \textit{region2vec} methods generate node neural embeddings based on attribute similarity, geographic adjacency and spatial interactions, and then extract network communities based on node embeddings using agglomerative clustering. The proposed GeoAI-based methods are compared with multiple baselines and perform the best when one wants to maximize node attribute similarity and spatial interaction intensity simultaneously within the spatial network communities. It is further applied in the shortage area delineation problem in public health and demonstrates its promise in regionalization problems. 

\textbf{Keywords:} 
GeoAI; community detection; spatial networks; graph convolutional networks; graph attention networks; neural network embeddings 

\end{abstract}

\maketitle


\section{Introduction}
Networks or graphs are often used to model many phenomena and relationships in the real world. For example, traffic moves along the transportation networks in geographic space, people connect through social networks, and academic scholars reference each other in the citation network. A network or graph usually consists of nodes and edges. The nodes can represent real-world entities and the edges are the connections between entities. In a spatial network, nodes can be geographic entities such as locations, road intersections or segments, and administrative areas \citep{zhu2020understanding}, and edges can represent the spatial interactions among nodes \citep{barthelemy2011spatial,gao2013discovering,liu2016incorporating,liu2019identifying,ye2021spatial}. The spatial interaction intensity can be used as the edge weights. Such spatial networks contain rich information and can be used to understand many geographic phenomena, for example, regionalization.

In the domain of geography, regionalization has always been of central importance. ``Understanding the idea of region and the process of regionalization is fundamental to being geographically informed" \citep[p.70]{geography1994geo}. The process of regionalization helps geographers and urban planners understand the complexity of many ongoing phenomena and gain insights into human behaviors and societies. In urban science, there is an increasing trend to study the functional cities in terms of location and spatial interactions with emerging big data and AI technologies~\citep{batty2021defining}.
Depending on the specific goals, the criteria used to divide regions can be multifaceted, including geographical, cultural, economic, political, etc. In the real world, many domains, such as health care, economics, or urban planning, require regions to be the basic unit for conducting further analyses, organizing and summarizing patterns. For example, comparing hospital visit statistics at the regional level can help policy-makers discover potential resource disparity in cities, make reasonable decisions, and propose strategies for future directions. Central to regionalization is the concept of spatial networks, which are structures that represent the complex relationships and interactions among different locations and regions. Spatial networks provide valuable insights into how different areas are interconnected and influence one another. However, they are always in a complex structure with many interrelated components that are hard to model and analyze.

Community detection algorithms have been widely used to extract information from complex networks and detect densely connected nodes in a network. Through the detected communities, studies can further identify the hidden relationship among network components and reveal the underlying network structures \citep{Su2021, He2018, expert2011uncovering}. The two major sources of information in a network (or a graph) used in community detection are the topological information represented by edges and the attribute information of nodes. Most of the traditional community detection algorithms partition the networks primarily based on topological structures. For example, the minimum-cut method partitions the graph by minimizing the number of edges between communities, the hierarchical clustering method uses the topological type of similarity between node pairs to group nodes, and the modularity maximization considers the concentration of edges within communities \citep{flake2004graph, johnson1967hierarchical, newman2006modularity}. 

However, the node attributes can also represent crucial characteristics such as people's demographic information or the region's economic status and should be incorporated into the regionalization process. How to better integrate both node attributes and network topology remains a challenge for traditional methods. In addition, another limitation of traditional community detection algorithms is that the computational cost increases dramatically with the expansion of networks \citep{Su2021}. Real-world networks may contain millions of nodes and edges with complex information and high-dimensional attributes. 

The past few years have witnessed significant developments in the adoption of deep learning methods to solve the community detection problem. Deep learning offers the following advantages compared with traditional methods: (1) it can convert the high-dimensional input into low-dimensional representations while maintaining the important structure information~\citep{hinton2006reducing,hamilton2017inductive} or convert low-dimensional input into high-dimensional representation to generate learning-friendly location encodings~\citep{mai2022review}; (2) it can integrate multiple types of information in various formats. Therefore, it has shown powerful performance on community detection tasks \citep{Su2021}.

The integration of multiple information sources in deep learning is especially helpful for spatial networks. While other networks may only have one type of edge connection, nodes in spatial networks can have an additional  relationship related to their geographic proximity. In many cases, the geographic places or regions may be spatially adjacent to each other. For example, if a study area is divided into grids, each grid can have multiple neighboring grids surrounding it. Therefore, given a spatial network, two graphs can be constructed: a graph for the spatial interactions among nodes where edges can be quantified using flow connections, and a graph for the geographic adjacency where edges can represent the binary relationship of whether two nodes are adjacent by sharing boundary or vertices. According to the first law of geography, ``near things are more related than distant things'' \citep{tobler1970computer}, which means that the nodes in spatial networks are naturally affected by their geographic neighbors, and such effects decay as two nodes become further away \citep{liu2012understanding}. The relationship can also be interpreted as a similarity measure, where nodes that are closer in geographic spaces tend to be more similar. 

The convolutional graph-based models are great candidates to model such relationships. Convolutional Neural Networks (CNNs) were first proposed for image analysis and can consider input topology by updating each input based on its surrounding inputs in a filter \citep{lecun1995convolutional}. Graph Convolutional Networks (GCNs) were further proposed for graph-structured data to apply CNN directly on graphs. Instead of using a fixed filter, the surrounding relationship is determined by the edge connections. In fact, GCN has gained great attention due to its good performance on node classification for graph-structured data \citep{He2018, Kipf2017}. GCN leverages the inherent structure of the data and aggregates neighbor nodes' information when updating the central node's representation \citep{Kipf2017}. Later on, Graph Attention Networks (GATs) were proposed to improve the neighborhood aggregation function in GCN, where a self-attention mechanism is used to learn the neighborhood importance for each pair of nodes \citep{velickovic2017graph}. Compared with GCN, which usually uses a fixed rule to aggregate neighborhood information, GAT uses the implicit attention coefficients to reflect that neighbors can have different importance to the central nodes. The advantages of GCN and GAT in modeling graph-structured data make them very useful for community detection on spatial networks and geographic contexts learning in GeoAI, especially for incorporating spatial concepts and structures such as spatial dependence, geo-semantics, and neighborhoods~\citep{zhu2020understanding, de2023graph,janowicz2020geoai,hu2024five}.

However, there are limited studies on applying convolutional-based neural networks to community detection tasks in spatial networks due to a few challenges. The lack of labels in community detection tasks makes it hard to train GCN or GAT models directly, as they are usually supervised or semi-supervised. How to solve the unsupervised learning problem using supervised models remains a challenge~\citep{xiao2022graph}. Second, although GCN and GAT are used to understand graph structures and generate latent representations for nodes, their goal is not community detection or regionalization-oriented \citep{Jin2019a, liang2022region2vec}. The learning objective for GCN and GAT needs to be re-designed to meet the requirements of community detection tasks.

With the consideration of those problems, the following research question (RQ) is asked:

\textit{RQ. How can node attributes and edge connections in spatial networks be combined simultaneously to identify regions  using GeoAI methods?}

To answer the research question, we designed a novel GAT-based unsupervised learning method for community detection and regionalization in this research, which extended the previous GCN-based learning method \citep{liang2022region2vec} to build a family of GeoAI-enhanced unsupervised learning methods that are guided by a community detection-oriented loss. The GeoAI-enhanced methods consider two types of edge relationships in the spatial network: spatial interactions and geographic adjacency, and combine multi-attribute information using the GCN and GAT models to learn node representations.
The communities (i.e., geographic regions) are further identified through an additional clustering step. The proposed GeoAI-enhanced community detection methods are called \textit{region2vec}; the name was firstly used by \cite{liang2022region2vec} to indicate a general type of methods that can extract latent representations (i.e., embeddings) based on regions' characteristics and spatial interactions. The used GCN and GAT models may be replaced with other graph neural network models based on different use cases. This study will focus on understanding the efficacy of the proposed methods in spatial network community detection.

The major contributions of this research are summarized as follows:

(1). We propose a family of GeoAI-based community detection algorithms that simultaneously consider both node attributes and edge connections using graph deep learning with geographic domain knowledge, i.e., the spatial dependence in Tobler’s First Law of Geography~\citep{tobler1970computer} and the similarity configuration in the Third Law of Geography~\citep{zhu2022third},  and overcomes the limitations of traditional community detection methods and clustering methods.

(2). We design a community-oriented loss that fills the gap in solving unsupervised learning tasks using (semi-)supervised models (GCNs and GATs) in geospatial regionalization tasks. The flexibility of inserting desirable node attributes and adopting available spatial interaction networks allows the broad applications of the proposed method.


The remainder of the sections are organized as follows: we first summarize the related literature on communication detection and graph embedding in Section 2, and we introduce the proposed community detection models in Section 3 and followed by the introduction to data and case study background in Section 4. We then present comparative analyses of the proposed method against other baselines and results in one public health application in Section 5. Finally, we conclude our research and share some directions for future work in Section 6.

\section{Related Work}
\subsection{Community Detection Algorithms on Spatial Networks}
As mentioned above, most of the traditional community detection algorithms are based on topological structures. Among them, the modularity-based maximization methods have particularly performed well on spatial networks \citep{gao2013discovering,expert2011uncovering}. The modularity proposed by \citet{newman2006modularity} is a measure of how good a graph partition is. It compares the number of edges that fall within the communities to a null network with edges placed randomly \citep{newman2006modularity}. A large modularity value indicates a robust community structure and dense connections among nodes within communities \citep{newman2006modularity, hu2018automated}. One of the most popular modularity maximization methods is Louvain \citep{blondel2008fast}. The Louvain method achieves the maximum modularity through two stages: (1) in a local network, move individual nodes to the community that maximizes the modularity gain, (2) aggregate the updated local networks and treat them as nodes recursively \citep{blondel2008fast}. The two stages are repeated until no improvement can be made. However, it has been found that the Louvain method tends to generate poorly-connected communities in some cases \citep{Wang2021}. The Leiden method was recently proposed to improve the Louvain method and guaranteed to generate well-connected communities \citep{traag2019louvain}.

The Louvain and Leiden methods have been applied to the public health domain to identify health service areas and spatially connected communities \citep{hu2018automated, Wang2021, pinheiro2020network, Wang2022}. The Louvain method was first applied to identify the hospital service areas based on the patient-to-hospital flows in Florida \citep{hu2018automated}. Later on, \citet{Wang2021} proposed spatially constrained Louvain and Leiden algorithms to identify health service areas. The algorithms have shown great performance in delineating health service areas with a goal of maximizing flows within communities and minimizing flows between communities \citep{hu2018automated, Wang2021}. They are effective and efficient for capturing the natural structure of the health visit patterns. In addition, traditional community detection methods have also been widely used in identifying spatial interaction communities using human mobility data, social media check-ins, cellphone call data, and community travel surveys \citep{ratti2010redrawing,liu2014uncovering,nelson2016economic,hou2021intracounty}. Besides the edge connections, the node attributes are also crucial to affecting community structures, such as people's demographic information and socioeconomic status, but they are not considered in those traditional edge-based community detection methods. 

\cite{liang2022region2vec} proposed a GCN-based unsupervised community detection method that considers both node attributes' similarity and edge connections. The algorithm demonstrates the potential of graph embedding in community detection tasks, but it cannot outperform the best baseline in every metric used. The effects of the spatial interaction are only used in the loss function, but they are not directly included in the GCN layers, which may also be one reason that the model does not perform the best. There are possibilities to improve the method and explore its applications in real-world examples of regionalization and community detection on spatial networks.

\subsection{Graph Embedding}

Graph embedding is a powerful representation learning technique that reduces the dimensionality of data by incorporating both node attributes and topological graph structures into vectors \citep{goyal2018graph}. With the information captured in vectors, node clustering is naturally extended from it. It has been shown in various scenarios that leveraging attribute  information in addition to the graph structure yields better results for node clustering \citep{cui2020adaptive}. Deep learning algorithms, including random walk-based algorithms and GCN-based algorithms, have both shown promising performance in graph embedding tasks \citep{goyal2018graph}.

Large-scale Information Network Embedding (LINE) was among the first graph embedding methods that scaled well with large networks \citep{grover2016node2vec, tang2015line}. LINE proposed to sample edges with probabilities based on weights and treat them as unweighted edges to solve the gradient explosion problem in stochastic gradient descent \citep{tang2015line}. Random walk-based algorithms such as DeepWalk and Node2vec, both inspired by the skip-gram models in natural language processing, aim to preserve high-order proximity between nodes in a graph by sampling fixed-length random walks and maximizing the probability of its neighbors in the walk \citep{perozzi2014deepwalk,grover2016node2vec}. Node2vec improves on DeepWalk by having a more flexible sampling strategy that balances between breadth-first-search and depth-first-search traversal, which allows it to learn a mixture of homophily and structural equivalence in a graph \citep{grover2016node2vec}. Another method called SDNE (Structural Deep Network Embedding) is also one of the earliest methods for graph embedding tasks \citep{wang2016structural}. It jointly optimizes for first-order and second-order proximities to preserve both the local and global structure of the graph \citep{wang2016structural}. This semi-supervised method is also robust to sparse networks \citep{wang2016structural}. The GraphSAGE framework later leveraged node features and aggregated attribute information in neighborhoods of the graph to improve node embeddings \citep{hamilton2017inductive}. 

In recent years, Graph Convolutional Networks (GCN) have gained tremendous attention for their strong capability in graph embedding and node representation learning, which are highly beneficial for further downstream prediction and clustering tasks \citep{zeng2019accurate, molokwu2020node}. Some of the latest GCN methods like GALA (Graph convolutional Autoencoder using LAplacian smoothing and sharpening), MAGNN (Metapath Aggregated Graph Neural Network), and AGE (Adaptive Graph Encoder) have all shown superior behaviors in popular node clustering datasets like Cora, IMDB, Wiki etc.\citep{fu2020magnn, cui2020adaptive, park2019symmetric}. Both GALA and AGE utilize graph encoder while MAGNN utilizes node attributes transformation and metapath aggregation \citep{fu2020magnn,cui2020adaptive,park2019symmetric}. Metapath is defined as “an ordered sequence of node types and edge types defined on the network schema, which describes a composite relation between the node types involved” \citep[p.2]{fu2020magnn}. MAGNN captures both semantic attributes and topological structures from neighboring nodes and metapath context in between \citep{fu2020magnn}. CommDGI (Community Deep Graph Infomax) developed a more specific GCN for community detection given the inherent unsupervised nature of community detection tasks compared to the other general-purpose graph representation learning problems by adding a clustering layer with community-oriented objectives like modularity \citep{zhang2020commdgi}. 

Graph Attention Networks (GATs) were further proposed to improve the neighborhood aggregation in GCN. In GCN, the weights between the neighbor nodes and the central node are usually explicitly defined, either through the structural properties of the graph or a learnable weight, and this can affect the generalization power of GCN. GAT defines the one-hop neighborhood weights implicitly through a self-attention mechanism \citep{velickovic2017graph}. The key idea is that different nodes in a neighborhood can have different importance to the central node. GAT has been proven to achieve or match state-of-the-art results and allows more interpretability through the learned attentional weights. Although GCN and GAT methods have been very successfully used in many tasks, their potential for community detection on spatial networks is still under-examined.

\section{Method}
We first introduce some definitions for the spatial network-based community detection problem with all the notations used in the following sections and briefly introduce the data sources used in the model. Then, the proposed \textit{region2vec} graph embedding algorithms and the clustering method for community detection are presented.

\subsection{Notations and Problem Definitions}
Graph $\bm{G} = (V, E)$ is defined via a set of nodes $V = (v_1, ..., v_n)$ (e.g., locations and regions), $|V| = n$ and edges $E$ with $e_{ij} = (v_i, v_j)$. $\bm{A} = [a_{ij}]_{n \times n}$ is an adjacency matrix representing geographic adjacency, where $a_{ij} = 1$ if $e_{ij} \in E$, otherwise $a_{ij} = 0$. $\bm{S} = [s_{ij}]_{n \times n}$ is a spatial interaction weight matrix, where $s_{ij}$ represents the flow intensity between node $v_i$ and $v_j$. An $ n \times m$ attribute matrix $\bm{X}$ is used to denote the node attributes.
 
 The community detection groups the $n$ nodes into $K$ communities $\{C_1, C_2, ..., C_K\}$ and each node will be assigned with a label $c_i$ indicating its community membership, $c_i \in \{1, 2, ..., K\}$.

\begin{figure*}[h]
  \includegraphics[width=\linewidth]{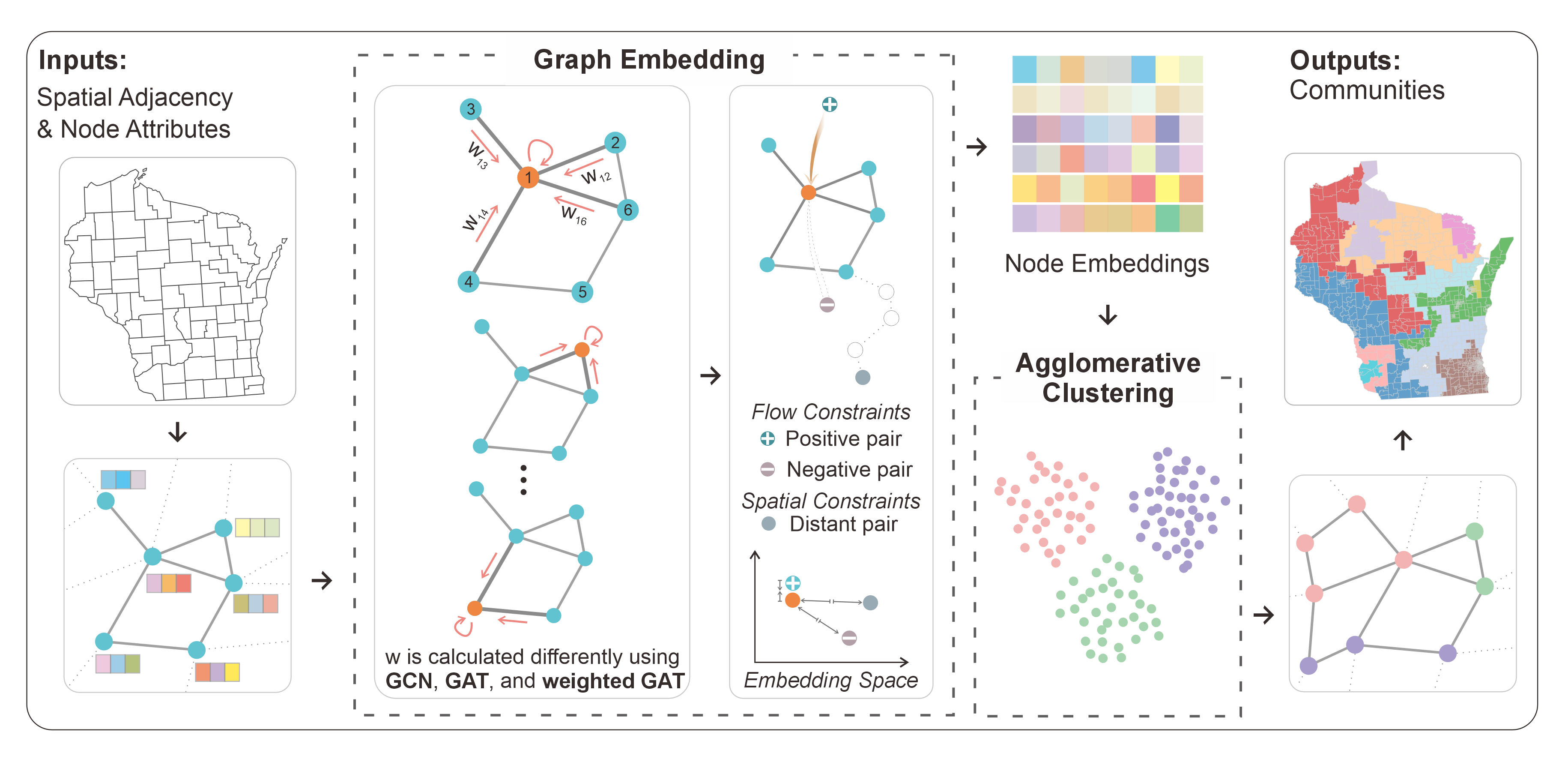}
  \caption{The workflow for community detection using the regions2vec method.}
  \label{fig:workflow}
\end{figure*}

\subsection{Algorithm}
Based on the proposed method, the identified communities should contain nodes (i.e., geographic regions) that satisfy the following three aspects: 1) share similar attributes, 2) have strong spatial interactions, and 3) are geographically adjacent. A two-stage community detection algorithm is proposed to fulfill the three requirements by considering node attributes and edge connections (spatial interactions and geographic adjacency) together. As shown in Figure \ref{fig:workflow}, stage one focuses on generating node representations encoded with the attribute, adjacency, and flow information, which enables the partition of network communities in the embedding space; Stage two focuses on clustering nodes into communities based on their similarities in the embedding space. 

\subsubsection{Stage One: Node Representation Learning}
Based on the first law of geography, nearby things are more related to distant things \citep{tobler1970computer}. Applying that to any node in a spatial network, its neighborhood nodes should have larger effects than distant nodes. Therefore, graph convolution becomes a natural tool that can aggregate neighbor information when updating the central node representation \citep{Kipf2017}. Two models are used respectively in the study to learn node embedding.

The first model used in the study is a GCN model with two convolutional layers. It was proposed by \cite{liang2022region2vec} and details about the model structure can be found in the original paper. 

 $Z^{(1)}$ and $Z^{(2)}$ are the outputs of the first and second graph convolutional layers, and $W^{(0)}\in \mathbb{R}^{m\times n_{hidden}}$ and $W^{(1)}\in \mathbb{R}^{n_{hidden}\times n_{output}}$ as the weights of two layers, the forward propagation model can be formalized as Equation \ref{eq:gc}:
 \begin{equation}
 \begin{aligned}
 Z^{(1)}&=ReLU (\tilde{D}^{-\frac{1}{2}}\tilde{A}\tilde{D}^{-\frac{1}{2}}X{{W}_{0}}),\\
 Z^{(2)}&= \tilde{D}^{-\frac{1}{2}}\tilde{A}\tilde{D}^{-\frac{1}{2}}Z^{(1)}{{W}_{1}},
 \end{aligned}
 \label{eq:gc}
 \end{equation}
 where $A$ and $I$ are the spatial adjacency and identity matrices, $\tilde{A}=A+I$, and $\tilde{D}$ is the degree matrix of $\tilde{A}$. 

The second model is the GAT model. It aggregates the neighborhood nodes with a self-attention mechanism. We followed the original GAT model proposed by \cite{velickovic2017graph}. The input is a set of node features, $\bold{z} = \{{\vec{z_1},\vec{z_2},..., \vec{z_n}}\}, \vec{z_i} \in \mathbb{R}^{m}$, where $n$ is the number of nodes, $m$ is the number of features in each node. The layer generates a new set of node representations, $\bold{z}^\prime = \{\vec{z_1^\prime},\vec{z_2^\prime},..., \vec{z_n^\prime}\}$, $\vec{z_i^\prime} \in \mathbb{R}^{m^\prime}$ (the new node representations may have a different cardinality $m^\prime$). 

A shared linear transformation weight matrix $W \in \mathbb{R}^{m^\prime \times m}$ is applied to every node. The self-attention mechanism $a:\mathbb{R}^{m^\prime}\times \mathbb{R}^{m^\prime} \mapsto \mathbb{R} $ then computes attention coefficients:
\begin{equation}
\centering
    e_{ij} = a(W\vec{z_i}, W\vec{z_j})
\end{equation}
For each node $i$, we only compute $e_{ij}$ for nodes $j \in N_i$, where $N_i$ is the neighborhood of node $i$. The softmax function is used to normalize the attention coefficients:
\begin{equation}
    \alpha_{ij} = \operatorname{softmax}_j(e_{ij}) = \frac{\operatorname{exp}(e_{ij})}{\sum_{k \in N_i}\operatorname{exp}(e_{ik})}
\end{equation}
The final output representation for every node is a linear combination of the features weighted by the normalized attention coefficients:
\begin{equation}
\centering
    \vec{z^\prime_i} = \sigma (\sum_{j \in N_i} \alpha_{ij}W\vec{z_j})
\label{eq:traditionalGAT}
\end{equation}

However, the attention coefficients are learned only from the node features in the original model based on a binary adjacency matrix. But GAT can not consider the case when there is a weighted input matrix that may contain different types of edges \citep{velickovic2017graph}. We added an additional weight $s_{ij}^{\prime}$ (a normalized flow intensity coefficient, $s_{ij}^{\prime} \in [0,1]$) to the attention coefficient to adjust the effects of spatial interaction on neighborhood node importance. The additional weight matrix $S^{\prime}$ is generated with a threshold $t^{\prime}$ to remove small-value flows, then the spatial flows are re-scaled and normalized to make sure the coefficients are comparable with the attention coefficients. We call this model the weighted GAT model:
\begin{equation}
\centering
    \vec{z^\prime_i} = \sigma (\sum_{j \in N_i} s^{\prime}_{ij} \alpha_{ij}W\vec{z_j})
\label{eq:weightedGAT}
\end{equation}

Both the traditional GAT following the equation \ref{eq:traditionalGAT} and the weighted GAT model using equation \ref{eq:weightedGAT} are adopted in the study. The input for GAT is the geographic adjacency matrix and the spatial positive flow matrix with a threshold $t$ as shown in Equation \ref{eq:GAT_input}. 

\begin{equation}
\centering
    A_{GAT} = A + S_{p_t}
\label{eq:GAT_input}
\end{equation}

However, GCN and GAT are not community detection-oriented models. We proposed a new loss function to guide the learning process. The loss function contains two constraints. The spatial interaction flow constraint will draw nodes with spatial interactions (positive pairs) closer and push nodes without spatial interactions (negative pairs) further in the embedding space. The spatial distance constraint uses a hop-distance threshold to discourage distant nodes from having similar embedding. The loss function is shown in Equation \ref{eq:loss}:
\begin{equation}
\begin{aligned}
L_{hops} &= \sum{\frac{\mathbb{I}(hop_{ij} > \epsilon)d_{ij}}{\log(hop_{ij})}},\\
L = &\frac{\sum_{p=1}^{N_{pos}}{\log(s_{p})d_{pos_{p}}}/N_{pos}}{\sum_{q=1}^{N_{neg}}{d_{neg_q}}/N_{neg} + L_{hops}},
\label{eq:loss}
\end{aligned}
\end{equation}
where $hop_{ij}$ is the number of hops of the shortest path between $v_i$ and $v_j$ in the graph, and $d_{ij}$ is the euclidean distance between the node embedding. $\mathbb{I}(\cdot)$ is set to 1 if $hop_{ij} > \epsilon$, or 0 otherwise. $pos_p, p\in [0,N_{pos}]$ and $neg_q, q\in [0,N_{neg}]$ represent the positive and negative pairs based on spatial interactions, where $N_{pos}$ and $N_{neg}$ are numbers of positive and negative pairs.  The pseudocode of \textit{region2vec} using the GCN model is shown in Algorithm \ref{alg:gcn} \citep{liang2022region2vec} and the pseudocode of \textit{region2vec} using the weighted GAT model is shown in Algorithm \ref{alg:gat}.

\RestyleAlgo{ruled}

\begin{algorithm}
	\caption{Region2Vec with GCN}\label{alg:gcn}
	\SetKwInput{KwInput}{Input}
	\SetKwInput{KwOutput}{Output}
	\KwInput{Graph $\bm{G} = (V, E)$; adjacency matrix $\bm{A}$; flow matrix $\bm{S}$; input features $\bm{X}$; the shortest path $hops_{i,j}, \forall i,j\in V$ and threshold $\epsilon$; number of layers $L$; weight matrices $W^l,\forall l \in \{0,\cdots,L-1\}$}
	\KwOutput{Node representations $\bm{z}_v$ for all $v\in V$}
	$Z^{(0)} \gets \bm{X}$\;
	$\tilde{A} \gets A+I$\;
	$pos_m \gets (i,j)$, for all $s_{ij} > 0$\;
	$neg_n \gets (i,k)$, for all $s_{ik} = 0$\;
	\For{each iter}{
		\For{$l=0,\cdots,L-1$}{
			$Z^{(l+1)}=ReLU (\tilde{D}^{-\frac{1}{2}}\tilde{A}\tilde{D}^{-\frac{1}{2}}Z^{(l)}{{W}^{l}})$\;
			
		}
		$d_{ij} = \|z_i-z_j\|$\;
		$L_{hops} = \sum{\mathbb{I}(hop_{ij} > \epsilon)d_{ij}/\log(hop_{ij})}$\;
		$L = \frac{1}{N_{pos}}\sum_{p=1}^{N_{pos}}{\log(s_{p})d_{pos_{p}}}/
		(\frac{1}{N_{neg}}\sum_{q=1}^{N_{neg}}{d_{pos_{q}}}+L_{hops})
		$\;
		Compute $g \gets \nabla L$\;
		Conduct Adam update using gradient estimator $g$
	}
	$\bm{z}_v \gets z_v^{L}, \forall v\in V$
\end{algorithm}

\begin{algorithm}
\caption{Region2Vec with weighted GAT}\label{alg:gat}
\SetKwInput{KwInput}{Input}
\SetKwInput{KwOutput}{Output}
\KwInput{Graph $\bm{G} = (V, E)$; adjacency matrix $\bm{A}$; flow matrix $\bm{S}$; positive flow matrix $\bm{{S_{p_t}}}$ with a threshold $t$; the normalized flow weight matrix $\bm{S^{\prime}}$ with a threshold $t^{\prime}$; input features $\bm{X}$; the shortest path $hops_{i,j}, \forall i,j\in V$ and threshold $\epsilon$; number of layers $L$; weight matrix $\bm{W}$}
\KwOutput{Node representations $\bm{z}_v$ for all $v\in V$}
$Z^{(0)} \gets \bm{X}$\;
$A_{GAT} \gets A+S_{p_t}$\; 
$pos_m \gets (i,j)$, for all $s_{ij} > t$\;
$neg_n \gets (i,k)$, for all $s_{ik} = 0$\;
\For{each iter}{
\For{$l=0,\cdots,L-1$}{
    $\vec{z^\prime_i}$ = $\sigma (\sum_{j \in N_i} s_{ij}^{\prime}\alpha_{ij}W\vec{z_j})$

}
$d_{ij} = \|z_i-z_j\|$\;
$L_{hops} = \sum{\mathbb{I}(hop_{ij} > \epsilon)d_{ij}/\log(hop_{ij})}$\;
$L = \frac{1}{N_{pos}}\sum_{p=1}^{N_{pos}}{\log(s_{p})d_{pos_{p}}}/
(\frac{1}{N_{neg}}\sum_{q=1}^{N_{neg}}{d_{pos_{q}}}+L_{hops})
$\;
Compute $g \gets \nabla L$\;
Conduct Adam update using gradient estimator $g$
}
$\bm{z}_v \gets z_v^{L}, \forall v\in V$
\end{algorithm}

\subsubsection{Stage two: Agglomerative Clustering}
The second stage is conducting clustering to obtain the final community memberships based on the node neural embeddings from the first stage. We used an agglomerative clustering method to aggregate similar nodes (i.e., regions) into larger groups. As a bottom-up approach, each node is an independent cluster at first and is grouped successively to form the final clusters \citep{scikit-learn}. We also enforce the connectivity constraint in the clustering algorithm to allow only clusters that are geographically adjacent to be merged together. Many regionalization problems using spatial networks might have an inherent spatial contiguity requirement to support further interpretation and analysis of the obtained regions. Therefore, it is necessary to have this spatial contiguity setting in the algorithm.

\subsection{Baseline Algorithms}
The following community detection algorithms are included in this study as baselines to compare with our proposed GeoAI-enhanced \textit{region2vec} models.

\subsubsection{Louvain community detection}
The Louvain algorithm \citep{blondel2008fast} first generates small communities through local modularity gain maximization, then the identified communities are treated as nodes and recursively aggregated again. The input of the Louvain method is the spatial flow network and it is used to detect communities using only human mobility flow connections in this study. 

\subsubsection{Leiden community detection}
The Leiden algorithm is a recent improvement on the Louvain algorithm. The Louvain algorithm has been found to generate arbitrarily badly connected communities \citep{traag2019louvain}. The Leiden added an additional partition refinement stage to ensure that communities are guaranteed to be well connected. The input of the Leiden method is also the spatial flow network.

\subsubsection{Random walk based model}
Two random walk based graph embedding models, Deepwalk \citep{perozzi2014deepwalk} and Node2vec \citep{grover2016node2vec} are used as baselines. Deepwalk uses short walks to generate random paths of connected nodes \citep{perozzi2014deepwalk}. The Node2vec algorithm further develops a biased random walk procedure to control different node neighborhood exploration strategies \citep{grover2016node2vec}. The input of the two methods is the spatial adjacency matrix. After obtaining the node embedding, the same agglomerative clustering algorithm is applied to generate the final communities.

\subsubsection{LINE}
Large-scale Information Network Embedding (LINE) proposed by \citet{tang2015line} is a network embedding method. It specifically preserves the local and global network structure through a carefully designed objective function that considers both first-order and second-order proximities. The input of LINE is the spatial adjacency matrix.

\subsubsection{K-Means}
The K-Means clustering algorithm can group nodes with similar attributes and assign each node to the cluster that is the nearest by measuring feature space distance \citep{macqueen1967some}. The K-Means clustering algorithm takes the node attributes as the only input and cannot consider the edge connections.

\subsection{Evaluation Metrics}
Four individual metrics and one synthetic metric are used to compare the community detection methods' performance from different perspectives. 

\subsubsection{Intra-Flow Ratio}
The intra-flow ratio is used to measure the ability to group nodes with strong flow connections. \cite{liang2022region2vec} proposed a similar metric called the intra-inter flow ratio. The intra-flow ratio used in this paper is a more robust metric with a fixed range of [0,1] that supports comparison across different scenarios. It divides the sum of edge weights within each community (intra-flow weights) by the sum of intra-flow and inter-flow edge weights in the whole spatial network. The equation for computing this ratio is shown in Equation \ref{eq:flow_ratio}, where $s_{ij}$ represents the flow intensity between nodes $i$ and $j$. This ratio ranges from 0 to 1, with a higher value indicating a better performance to group nodes with strong connections together.

\begin{equation}
\begin{gathered}
      R_{Intra Flow} = \frac{\sum_{c_i=c_j} s_{ij}}{\sum s_{ij}}; \\
    c_i, c_j\in {1, 2,\cdots, K} \ \text{(}K \textrm{ is the number of communities)}
\end{gathered}
\label{eq:flow_ratio}
\end{equation}


\subsubsection{Inequality}

The inequality metric was initially designed to reflect the infrastructure inequality \citep{pandey2022infrastructure}. The inequality reflects the heterogeneity of all the samples and how it varies with the average value. It is calculated using Equation \ref{eq:inequality} where $\sigma$ is the standard deviation and $\mu$ is the mean. In this study, the inequality is measured using all the node features. For each feature, the inequality in each community is calculated and the median inequality in all communities is selected as a representation for that feature dimension. The final score is the product of all inequality scores (Equation \ref{eq:inequality_product}) where $m$ is the number of features. The inequality is then converted into a range of 0 to 1 using the min-max normalization (Equation \ref{eq:min_max}). A value of 1 means maximum inequality and a value of 0 means minimum inequality (i.e., a better performance).s

\begin{equation}
\begin{aligned}
    I_i = \frac{\sigma_i}{\sqrt{\mu_i (1-\mu_i)}}; 0 < \mu_i < 1
\label{eq:inequality}
\end{aligned}
\end{equation}

\begin{equation}
\begin{aligned}
    I = \prod_{i=1}^{m} I_i
\label{eq:inequality_product}
\end{aligned}
\end{equation}

\begin{equation}
    I_{norm} = \frac{I - I_{min}}{I_{max} - I_{min}}
\label{eq:min_max}
\end{equation}

\subsubsection{Cosine Similarity} We use cosine similarity to measure the ability of grouping nodes when they share similar attributes. Cosine similarity calculates the L2-normalized dot product of vectors \citep{scikit-learn}, which is the cosine of the angle between the two vectors, and therefore, is not dependent on the magnitudes of the vectors \citep{manning2010introduction}. The cosine similarity for all pairwise nodes in each community and the median values in all communities are calculated.

\subsubsection{Synthetic Score}
The previous three metrics are measured along different dimensions of the community detection result. To summarize all the different aspects, a synthetic score $S$ is calculated as the product of the three metrics. To make it comparable, we need to make sure all metrics are within the same range. For intra-flow ratio and cosine similarity, they range from 0 to 1 with a monotonic increase to indicate a better performance. For inequality, originally, it ranges from 0 to infinity,  after the min-max normalization, it ranges from 0 to 1 with a smaller value indicating a better performance. Therefore, it is then converted using a monotonically decreasing function to match with the other two metrics. The synthetic score $S$ is then calculated based on Equation \ref{eq:synthetic}, where a larger synthetic score represents a better performance.

\begin{equation}
\begin{aligned}
    S = R_{Intra Flow} * Cosine Similarity * (1 - I_{norm})
\label{eq:synthetic}
\end{aligned}
\end{equation}

\subsubsection{Join count ratio}

The join count ratio is used to measure the spatial dependence \citep{cliff1973spatial,kang2022sticc}. It is based on the join count statistics to measure the proportion of neighborhood nodes that belong to the same community across all the neighborhood nodes. For each pair of geographically adjacent nodes that share a boundary, whether their community is the same will be calculated. The join count ratio is shown in Equation \ref{eq:join_count_r}, where $J_{same}$ is the number of neighborhood pairs that belong to the same community and $J_{diff}$ is the number of neighborhood pairs belongs to the different communities. The $J_{ratio}$ ranges from 0 to 1 and represents how well geographic-adjacent nodes are grouped into the same community. A higher value indicates a stronger ability to maintain spatial contiguity.

\begin{equation}
\begin{aligned}
    J_{ratio} = \frac{J_{same}}{J_{same} + J_{diff}}
\label{eq:join_count_r}
\end{aligned}
\end{equation}

\section{Data and Case Study}

\subsection{Data}
The spatial network is constructed using the SafeGraph business venue database \citep{SafeGraph}. The visit patterns to different places are collected from anonymous smartphone users and each visit contains an origin location and a destination visited place. The place visit origins and destinations are then aggregated to the same geographic level (census tract in this study) to build the spatial interaction network \citep{kang2020multiscale}. Note that other geographic scales of regions, such as census block groups and counties, can also be used to construct spatial interaction networks. The shapefile of U.S. census tracts is used to construct the geographic adjacency matrix \citep{bureau2023tiger}. 
The census tract level demographic attributes (e.g., income, population, race) are gathered from U.S. Census American Community Survey (ACS).

\subsection{Health Service Area Case Study}
One application of the proposed GeoAI-enhanced community detection methods in spatial networks is to support the Rational Service Area (RSA) development, which is a critical issue in the public health domain. The rational service areas are self-contained geographic units that reflect how people move and seek health services, and they should be defined based on travel patterns, physical barriers or social-economic similarities \citep{hrsa2020, lopes2000state}. Therefore, the developed RSA designation should reflect how people move across the region and also the social-economic factors among local residents, which requires our proposed \textit{region2vec} method considering both node features and spatial interactions. However, existing rational service areas are mainly developed by manual work with local health knowledge \citep{hu2018automated}. There is a lack of systematic approaches to formalize the rational service area development and support adaptive, repeated and flexible designation updates. 

The rational service areas serve as the basic geographic unit for identifying Health Professional Shortage Areas (HPSAs). After establishing rational service areas, each of them is evaluated from multiple spatial and non-spatial aspects to generate a health shortage score. The four scoring criteria are the population-to-provider ratio, the percentage of the population at 100\% federal poverty level, infant health index (based on infant mortality rate or low birthweight rate), and travel time or distance to the nearest source of non-designated provider \citep{hrsa2020}. Areas with a score larger than zero are defined as the Health Professional Shortage Area and will receive additional funding to support local health providers. The ultimate goal of RSA development is to accurately group areas that lack health services so that they can be identified in the HPSAs scoring process.

The communities detected from the proposed \textit{region2vec} method and several baselines are used as the delineated rational service areas, respectively. The generated communities are the input of the HPSAs scoring process and the results are then evaluated based on the following metrics. The number of delineated HPSAs, the area covered, and the population-to-provider ratio will be calculated. The provider locations are downloaded from the Bureau of Health Workforce \citep{hrsa2023bhw}. The population-to-provider ratio can reflect, on average, how many people are assigned to one provider. The higher this value is, the more severe the health shortage this area is facing.

\section{Results}

There are two major parts of the results section. First, the performance of community detection on spatial networks measured by various metrics will be compared together with visualizations for all the introduced methods. Then, a case study for how the proposed \textit{region2vec} method can be applied to the rational service area development in public health is presented. The Louvain and Leiden algorithms generate a different number of optimal communities based on the input network; the Louvain algorithm identifies 14 communities, and the Leiden algorithm identifies 12 communities. The number of communities for the rest of the methods which require this predefined parameter is set as 14. The performance of different community number settings is further discussed in section \ref{sec:sensitivity}. For the traditional GAT model, the threshold $t$ in $S_{p_t}$ is set to 5 to include most of the positive flow edges. For the weighted GAT model, the threshold $t$ in $S_{p_t}$ is set to 200 and the threshold $t^{\prime}$ in $S^{\prime}$ is set to 100.


\begin{table}[h]
\centering
\caption{The metrics comparison of all methods. (In \textbf{bold}:
best; \underline{Underline}: second best)}
\label{tab:metrics}
\begin{tabular}{cclcl}
\hline
Methods & \begin{tabular}[c]{@{}c@{}}Intra-Flow \\ Ratio\end{tabular} & \multicolumn{1}{c}{\begin{tabular}[c]{@{}c@{}}Normalized \\ Inequality\end{tabular}} & \begin{tabular}[c]{@{}c@{}}Cosine\\ Similarity\end{tabular} & \multicolumn{1}{c}{\begin{tabular}[c]{@{}c@{}}Synthetic \\ Score\end{tabular}} \\ \hline
Region2vec-weightedGAT & \textbf{0.843} & {\ul 1.879E-07} & {\ul 0.975} & \textbf{0.821} \\
Region2vec-GAT & 0.702 & 1.523E-04 & 0.967 & 0.679 \\
Region2vec-GCN & 0.782 & 1.130E-05 & 0.974 & 0.762 \\
DeepWalk & 0.721 & 6.469E-04 & 0.960 & 0.691 \\
LINE & 0.214 & 1.000E+00 & 0.872 & 0.000 \\
Node2vec & 0.731 & 1.394E-03 & 0.951 & 0.694 \\
K-Means & 0.305 & \textbf{0.000E+00} & \textbf{0.983} & 0.299 \\
Louvain & 0.829 & 1.711E-04 & 0.967 & {\ul 0.801} \\
Leiden & {\ul 0.831} & 2.879E-04 & 0.964 & {\ul 0.801} \\ \hline
\end{tabular}
\end{table}

\subsection{Comparison with the existing methods}
\subsubsection{Metric Comparison}
The results of different metrics used to compare method performances are listed in Table \ref{tab:metrics}. Overall, the proposed \textit{region2vec} method with weighted GAT has the best performance for the intra-flow ratio and the synthetic score and the second-best performance for inequality and cosine similarity. While K-Means is the best one for inequality and cosine similarity, it has the second lowest intra-flow ratio, which indicates that it fails to consider edge interaction during the clustering process compared with other methods. The proposed \textit{region2vec} method with weighted GAT is the best when one takes all aspects into consideration and this is also indicated by the synthetic score. 

First, the intra-flow ratio represents the ratio of intra-community flows out of all the flows. Our proposed \textit{region2vec} method with weighted GAT has the highest score, meaning that it performs the best in grouping nodes while considering the spatial interactions among them. The Leiden method and Louvain method take the spatial interaction matrix as the input and have similar results - they have the second-highest and third-highest flow ratio values respectively. The proposed \textit{region2vec} method with GCN has the fourth-best performance. After them, Node2vec and Deepwalk obtain similar ratios and are listed as fifth and sixth best. The \textit{region2vec} method with traditional GAT does not reach a very high intra-flow ratio. The K-Means method and LINE have the lowest ratios, much smaller than the rest of the methods.

A lower normalized inequality value means that nodes within communities have more similar attributes and are more equal. The K-Means clustering method obtains the lowest value. As K-Means groups nodes only using their features, it is expected to perform well in this metric. The proposed \textit{region2vec} method with weighted GAT has the second lowest normalized inequality, meaning that the nodes within each community are also very homogeneous in the feature space. The proposed \textit{region2vec} method with GCN has the third lowest normalized inequality, showing that, in general, our proposed method has advantages over other baselines in grouping nodes with similar attributes.

The cosine similarity is another metric to measure node attribute similarity in communities. K-Means clustering obtains the highest cosine similarity. The \textit{region2vec} method with weighted GAT is the second best and the \textit{region2vec} method with GCN is the third best. This demonstrates the ability to group nodes with similar attributes from another perspective, and the \textit{region2vec} methods outperform most of the baselines besides K-Means.

For the synthetic score, the \textit{region2vec} method with weighted GAT has the highest value, meaning that after combining all aspects, it is the best one with a comprehensive consideration across different dimensions, including node attributes and spatial interactions. The Louvain and Leiden methods have the second-best synthetic score, partly due to their good performance on the intra-flow ratio metric. The \textit{region2vec} method with GCN is the fourth best and also demonstrates the advantage of the proposed method. In summary, the proposed \textit{region2vec} method with weighted GAT has the best score and is in the top two across all the metrics used, showing that it can maintain a great balance between grouping nodes with strong spatial interaction and grouping nodes with similar attributes.

\begin{figure}[h]
\centering
  \includegraphics[width=0.9\linewidth]{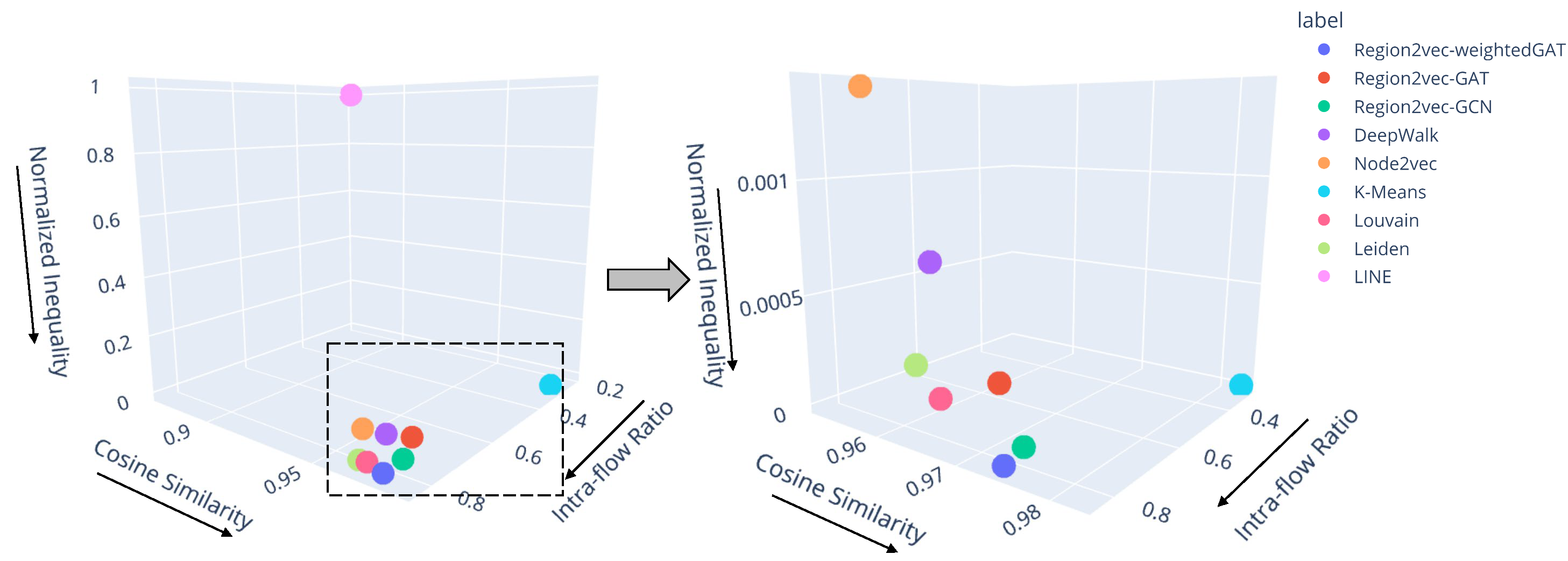}
  \caption{The 3D view of three metrics for all the methods. Left: the view of all methods, Right: the zoomed-in view without LINE}
  \label{fig:3d_metrics}
\end{figure}

In addition, to better understand the metrics, 3D plots are generated in Figure \ref{fig:3d_metrics}. The three dimensions are intra-flow ratio, cosine similarity, and normalized inequality, where the direction of arrows represents a better performance. On the left side, all methods are plotted where LINE is very distant from all other methods due to its large normalized inequality and low intra-flow ratio. On the right side, LINE is excluded to enable a zoomed-in view of other methods. It is clear to see that K-Means is away from others due to its low intra-flow ratio. DeepWalk and Node2vec have very high normalized inequality scores that separate them from other methods. The rest of the methods are clustered around the area with the best performances across three dimensions as the arrows indicate. Among them, \textit{region2vec} with weighted GAT and \textit{region2vec} with GCN are closer to the best-performance corner, showing the advantages of the proposed method.

\subsubsection{Sensitivity Analysis}\label{sec:sensitivity}

\begin{figure}[h]
	\centering
	\includegraphics[width=0.6\linewidth]{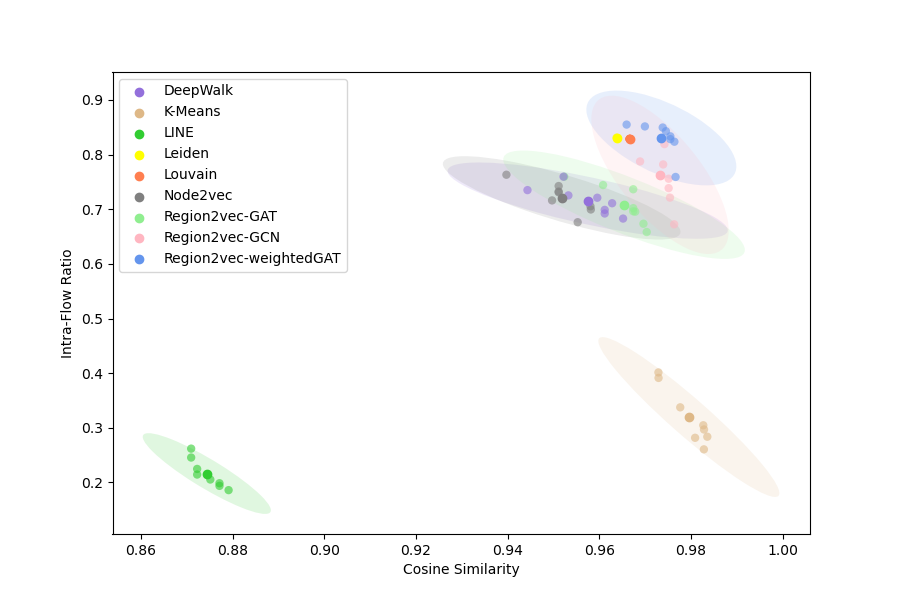}
	\caption{The distribution of all methods based on cosine similarity and intra-flow ratio.}
	\label{fig:cos_ratio}
\end{figure}

Since the ground-truth community structure for unsupervised machine learning is unavailable, there is a lack of knowledge of the exact number of communities. In the previous analysis, we set the number according to the Louvain algorithm for easy comparison. Here, to further verify the effectiveness of the proposed \textit{region2vec} methods, we conduct a sensitivity analysis to test the performance under different settings of community numbers. The result is shown in Figure \ref{fig:cos_ratio}, which demonstrates the scoring distribution of all methods based on cosine similarity and intra-flow ratio along with their scoring 95\% confidence-level standard deviation ellipse. For the same algorithm, dots with lighter colors represent results derived from different community number settings, while the darker one denotes the average score. As we can see, along the x-axis (cosine similarity), K-Means, the proposed \textit{region2vec} with weighted GAT and \textit{region2vec} with GCN have the highest values and can always outperform the rest of the methods. DeepWalk, Node2vec and \textit{region2vec} with GAT fall within a similar range of cosine similarities. LINE has the worst performance. Along the intra-flow ratio dimension, the proposed \textit{region2vec} with weighted GAT outperforms all other methods with very stable intra-flow ratios. The Louvain and Leiden methods are the second-best and third-best, and they are followed by \textit{region2vec} with GCN with some fluctuations in the ratio values. DeepWalk, Node2vec and \textit{region2vec} with GAT show similar ranges of intra-flow ratios and LINE is again the worst. In summary, the \textit{region2vec} with weighted GAT has very stable performances across different community numbers. Out of all the methods, it maintains a great balance of grouping nodes with similar attributes (indicated by cosine similarity) and grouping nodes with stronger flow connections (represented by the intra-spatial interaction flow ratio).

\subsubsection{Community Visualization}
%
%


\begin{figure}[H]
	\centering
	\includegraphics[width=0.99\linewidth]{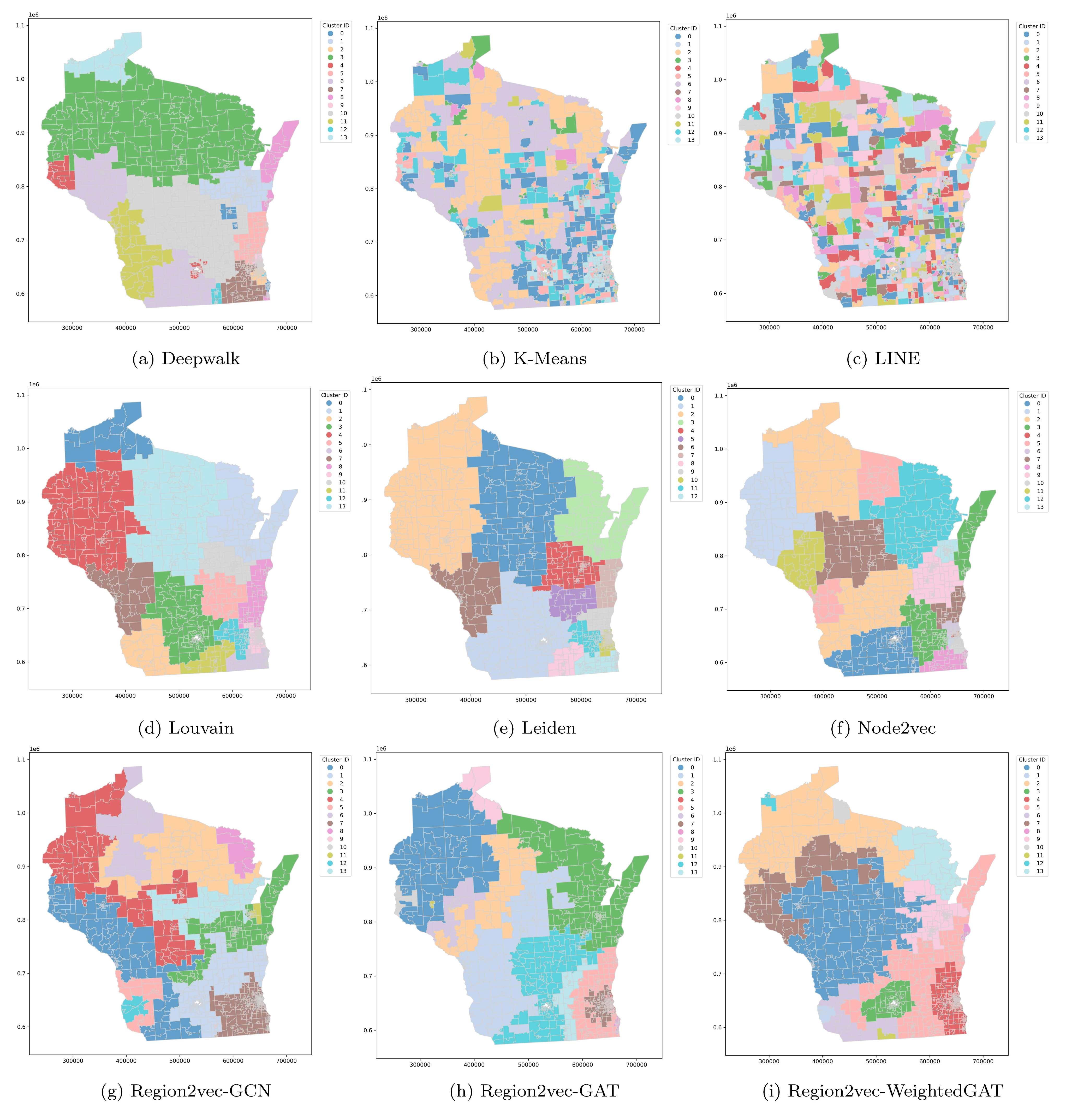}
	\caption{The resulting communities maps of all methods.}
	\label{fig:maps_all}
\end{figure}

Since the nodes in spatial networks are geographic regions (i.e., census tracts in this study), the resulting communities can be visualized on the map. The map visualization of the detected communities for all the methods is shown in Figure \ref{fig:maps_all}. The whole study area is divided into 14 communities except for Leiden (which identifies 12 communities), represented by different colors. Overall, \textit{region2vec} with GCN, GAT, weighted GAT, Deepwalk, Node2vec, Leiden and Louvain generate relatively regular shapes of communities that are mostly connected, while K-Means clustering and LINE have very scattered results. As the resulting spatial network communities should be spatially contiguous for most of the downstream tasks, the results from K-Means clustering and LINE cannot satisfy such requirements and are excluded from the following comparisons. Due to algorithm limitations, Deepwalk and Node2vec generate some communities that are actually spatial discontiguous, such as community 6 (in light purple) in Deepwalk and community 2 (in light orange) in Node2vec, which might affect the downstream tasks as some nodes in the same community are not geographically adjacent to each other.
To quantitatively evaluate how well the spatial contiguity of nodes is maintained, the join count ratio is calculated for each method and the result is shown in Table \ref{tab:join_count}. The Leiden method has the greatest ratio, meaning that it has the highest proportion of geographic neighborhood pairs that share the same community labels. The Louvain method has the second-highest ratio and is very similar to the Leiden method. The \textit{region2vec} with weighted GAT has the third highest ratio (above 0.9), indicating that the spatial contiguity is also well maintained. Similar to the visualization maps, LINE and K-Means have the lowest join count ratios and cannot maintain spatial contiguity. 

\begin{table}[h]
\centering
\caption{The join count ratios of all methods (In \textbf{bold}:
best; \underline{Underline}: second best).}
\label{tab:join_count}
\begin{tabular}{cc}
\hline
Methods                & Join Count Ratio     \\ \hline
DeepWalk               & 0.902                \\
K-Means                & 0.315                \\
LINE                   & 0.112                \\
Louvain                & \underline{0.929}     
\\     Leiden & \textbf{0.932}     
\\
Node2vec               & 0.900                \\
Region2vec-GCN         & 0.873                \\
Region2vec-GAT         & 0.838                \\
Region2vec-WeightedGAT & 0.904 \\ \hline
\end{tabular}
\end{table}

In terms of the area shape, Leiden, Louvain and Node2vec have more compact area shapes by visual examination. It is noticeable that the shape of communities in \textit{region2vec} tends to be particularly long or irregular. The geographic distances between census tracts in the same community may be very large due to the long shape of the community. One potential reason behind this is that there exists long-distance mobility flows in the spatial networks: people living in rural areas of Wisconsin need to travel long-distance to visit other areas and such connections may not be reflected in results from Leiden, Louvain, Deepwalk and Node2vec. This is one advantage of the proposed \textit{region2vec} algorithm: not only the areas that are geographically adjacent should be in the same community, but also areas that are connected by the transportation infrastructure (reflected in the daily human mobility flows). Therefore, the results of \textit{region2vec} may be more helpful for revealing regional human movement patterns such as commuting while maintaining the socioeconomic similarities of clustered regions. In particular, \textit{region2vec} with weighted GAT identifies two major urban areas in the State of Wisconsin, community 8 (in green) containing Dane County, where the state capital is located, and community 7 (in red) containing Milwaukee area - the largest city in Wisconsin.

\subsection{Case Study in Public Health}


\begin{figure}[h]
	\includegraphics[width=\linewidth]{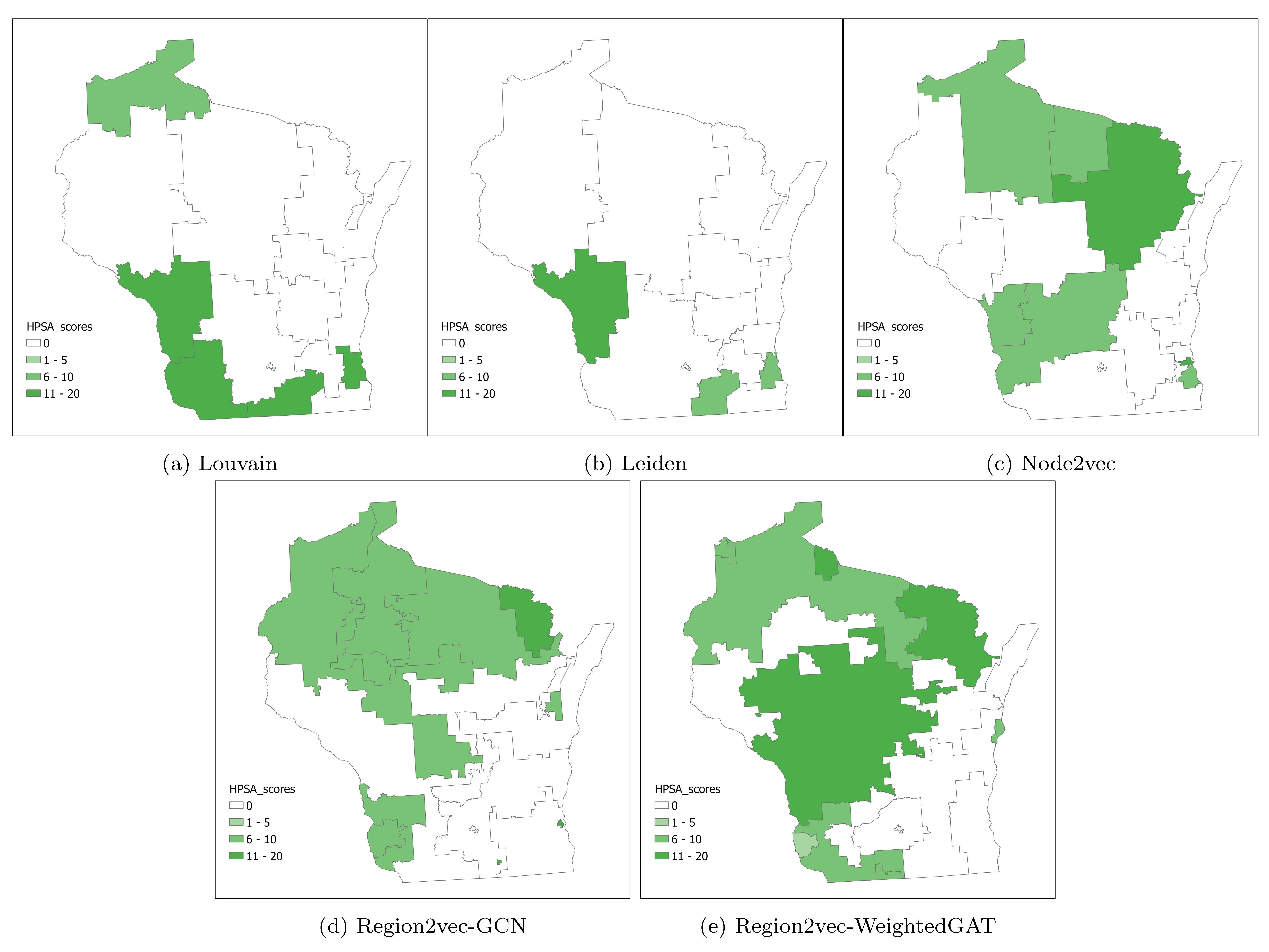}
	\caption{The final Health Professional Shortage Area with scores for three methods.}
	\label{fig:hpsa}
\end{figure}

\begin{table}[h]
\centering
\caption{The HPSA delineation performance of four methods (Bold indicates the best).}
\label{tab:hpsa}
\begin{tabular}{ccccccc}
\hline
Methods                & \begin{tabular}[c]{@{}c@{}}HPSA \\ number\end{tabular}& Population:Provider  & \begin{tabular}[c]{@{}c@{}}Total \\ Areas ($km^2$)\end{tabular} \\ \hline
Louvain                & 5                              & 4796.608  & 33728.846  \\ Leiden                & 3                              & 4926.709  & 14169.048  \\
Node2vec               & 5 & 5382.736  & 70718.458 \\
region2vec-GCN         & \textbf{9}&  \textbf{5900.010} &  75677.092  \\
region2vec-weightedGAT & \textbf{9} & 4779.635    & \textbf{89848.260}\\ \hline
\end{tabular}
\end{table}

Based on the community map visualization, the results of \textit{region2vec} with GCN, \textit{region2vec} with weighted GAT, Node2vec, Leiden and Louvain are selected to be applied in the Rational Service Area development problem in public health. The communities are used as the basic units, and then four scoring criteria are applied to find the shortage area. The final Health Professional Shortage Area (HPSA) scores for primary care are shown in Figure \ref{fig:hpsa}. The areas with green color are identified as Shortage Areas, and the transparency represents the value of the scores where lighter greenness has lower scores. As shown in Table \ref{tab:hpsa}, the \textit{region2vec} - weighted GAT method identifies a significantly large area coverage while Leiden and Louvain show much smaller area coverages. The \textit{region2vec} with GCN and Node2vec have similar area coverages. For HPSA numbers, the \textit{region2vec} with GCN and \textit{region2vec}-weighted GAT method generate the largest number of HPSAs (9), while Louvain and Node2vec have 5 HPSAs, and Leiden has 3 HPSAs. The number of HPSAs is an important criterion for measuring public health needs. The lower number of HPSAs in Leiden, Louvain, and Node2vec indicates that they may fail to identify some areas with potential health shortages. The population-to-provider ratio indicates the level of shortages as to how many populations are allocated to one provider. The \textit{region2vec}-GCN achieves the highest population-to-health provider ratios, meaning that it identifies the areas with the greatest shortage while  the \textit{region2vec} with weighted GAT maximizes the total covered areas. 

Based on the HPSAs identified by the proposed methods (Figure \ref{fig:hpsa_gcn}, and Figure \ref{fig:hpsa_gat}), it is found that most of northern Wisconsin is identified as Shortage Areas. These are also the rural areas in the state where there are fewer providers due to the large travel distances and sparser population. The shortage has been realized and confirmed through discussions with the state health officials, and it also demonstrates the potential of the proposed method to accurately identify health shortage areas. Compared with the manual process that most health service officials are using, the proposed \textit{region2vec} method offers a systematic approach that can identify the health service areas in a quantitative, repeatable, and adaptive manner. Furthermore, our approach offers the ability to insert new node features and use other available spatial networks. In case there are other health-related features, such as the number of residents with certain types of health conditions, the number of hospitals, and the number of providers in primary care, mental care, or dental care, they can be added to the algorithm to support more specific regionalization tasks. Similarly, the type of spatial networks can also be the hospital visit flow from patients to each hospital or the commuting flow patterns.

In summary, the proposed \textit{region2vec} methods are specifically designed for spatial networks that consider both the node attributes and spatial interactions, so the derived communities will have similar characteristics and strong spatial flow connections. As the designation of Rational Service Areas requires evidence of travel patterns or similar socio-economic attributes, the proposed GeoAI-enhanced community detection method on spatial networks meets the requirement and is very suitable to be applied to this problem. Through the comparisons with other baselines, the \textit{region2vec} methods family has shown its great performance using various metrics in practical applications.


\section{Conclusion}
This study proposed a new family of GeoAI-enhanced unsupervised community detection methods on spatial networks called \textit{region2vec}. Based on graph neural network models including GCN and GAT, the \textit{region2vec} methods use graph embedding to learn node representations with the consideration of multiple edge relationships and node attributes. The learning process is guided with a self-designed community detection-oriented loss so that nodes with strong spatial interactions, similar attributes, and geographic adjacency are encouraged to have similar representations in the embedding space. The communities (i.e., geographic regions) are further formed through a post-clustering process. By comparing with multiple baselines using different metrics, the GeoAI-enhanced proposed methods have shown the greatest performances in the flow intensity score and the combined synthetic score. In particular, the \textit{region2vec} method with weighted GAT integrates an additional spatial interaction weight on top of the attention coefficients and performs the best among all the methods. 

The \textit{region2vec} methods have also been applied to the rational service area development problem in public health and show their promise in solving regionalization problems using spatial networks, with the advantage of combining both edge connections and node attributes in the process. 
The proposed methods can also benefit several practical applications. For any regionalization tasks that involve inter-region connections or regional characteristics, one may apply the proposed \textit{region2vec} methods using the node attributes and the edges that are most appropriate. For example, some potential applications of \textit{region2vec} include political redistricting,  geodemographic classification, urban functional zone development, traffic zone delineation, etc. For political redistricting, it refers to the process of drawing electoral district boundaries for representative voting purposes. Recent studies have witnessed the difficulty of balancing multiple criteria in district planning and involving the ever-changing movement patterns. The proposed \textit{region2vec} can be a great candidate to consider both demographic/political factors and inter-district spatial interactions to identify communities of interest in the redistricting process~\citep{kruse2023bringing}. In this case, one may want to use all the node attributes related to redistricting and select the human mobility visits to certain types of places. The detected communities can be viewed as electoral districts. Similarly, for other studies, such as traffic zone division, one has the freedom to select certain node attributes from the transportation domain and use movement trajectory networks to represent the interactions. 

One future work of the proposed \textit{region2vec} can be expanding the original model to multiplex graphs using multiplex graph representation learning approaches~\citep{bielak2024representation}. In a multiplex graph, nodes can be connected by multiple types of relationships. For example, in a geographic network, geographic units can be connected by mobility flows or geographic adjacency relationships. In a social network, people can be connected by their friendships or by their hobbies. In case the edge relationships are within the same category or have similar value ranges, it is possible to design a weighted multiplex graph where the edge is a weighted sum/average of multiple relationships or using other fusion functions. The weight of each relationship can be tested and adjusted based on the actual scenarios. The weighted multiplex graph can be used as the input of the proposed method directly.  However, in many other cases, the relationships are not directly comparable. For example, in a geographic network, the geographic adjacency relationship of two geographic units is often binary, where 0 means not adjacent and 1 means adjacent. While the spatial interaction relationship (mobility flows in this study) between two nodes can range from 0 to infinity. How to find a balance to combine the two relationships based on the specific application can be challenging. 
Therefore, how to appropriately model the multiple relationships in the proposed method needs further exploration in future work.

Another future direction to explore is taking the overall community structures into consideration. It has been discovered that many geospatial networks have a scale-free property \citep{jiang2007topological, ma2020exploring}. It refers to the concept that in some networks, the distribution of node connections (or other features) would follow a power law, where only a few nodes would have many connections, and the rest of the nodes would have relatively few connections \citep{albert2002statistical}. Taking this concept to the communities, it means that there may exist far more small communities than large ones in complex networks and it has been proved by empirical evidence on the detected communities \citep{jiang2015defining}. Knowing this characteristic of the underlying communities would help better understand the community structure and design a better objective function when conducting community detection tasks using GeoAI models. Following the scale-free property, one may further learn the topological representation of the community as a whole on top of the geographic representations~\citep{jiang2019topological}. The topology allows the understanding of underlying spatial heterogeneity and scaling hierarchy of nodes in complex networks. For example, \cite{jiang2019geographic} established living structures at different levels of scale in a nested manner based on millions of street nodes. Future community detection methods may further explore improvements by better utilizing such topological structures.

There are some other aspects that need further improvement. The shapes of generated communities using \textit{region2vec} are not compact enough compared with other methods such as Node2vec or Louvain as the current spatial constraints cannot strictly affect the shape of communities. Our future work will aim to develop different approaches of integrating multiple graphs and node attributes to generate more compact communities. 
Also, more node attributes will be included in the algorithm to help generate communities with more similar regional characteristics. Besides the healthcare application, we also plan to add more experiments in different application domains using other types of data to further test the model generalizability.

This research shows the ability to use graph embedding on spatial networks to solve regionalization problems. There are interesting relationships in spatial networks that can be further explored using deep learning models, such as geographic contiguity requirements and certain area shape requirements (concave or convex). This study contributes to the increasing interest in GeoAI development in GIScience, urban analytics, and beyond \citep{grekousis2019artificial,janowicz2020geoai,liu2022review,de2023geoai}.  

\section*{Code Availability}
The codes that support the findings of this study are available at the following GitHub repository: \url{https://github.com/GeoDS/region2vec-GAT/}

\section*{Acknowledgement}
We acknowledge the funding support from the County Health Rankings and Roadmaps program of the University of Wisconsin Population Health Institute, Wisconsin Department of Health Services, and the National Science Foundation funded AI institute [Grant No. 2112606] for Intelligent Cyberinfrastructure with Computational Learning in the Environment (ICICLE). Any opinions, findings, and conclusions or recommendations expressed in this material are those of the author(s) and do not necessarily reflect the views of the funders.

\bibliographystyle{apalike}
\bibliography{references}

\end{document}